# Enhanced Population on Ionic Excited States by Synchronized Ionization and Multiphoton Resonance

Yewei Chen[1,3,#], Hongbin Lei[2,#], Quanjun Wang[1], Hongqiang Xie[4], He Zhang[1], Xu Lu[1,5], Ning Zhang[1,5], Shunlin Huang[1], Yuzhu Wu[1,6], Jianpeng Liu[2], Qian Zhang[2], Yi Liu[3], Zengxiu Zhao[2], Jing Zhao[2,†], Jinping Yao[1,*]

[1]*State Key Laboratory of High Field Laser Physics, Shanghai Institute of Optics and Fine Mechanics, Chinese Academy of Sciences, Shanghai 201800, China*

[2]*Department of Physics, National University of Defense Technology, Changsha 410073, China*

[3]*School of Optical-Electrical and Computer Engineering, University of Shanghai for Science and Technology, Shanghai 200093, China*

[4]*School of Science, East China University of Technology, Nanchang 330013, China*

[5]*Center of Materials Science and Optoelectronics Engineering, University of Chinese Academy of Sciences, Beijing 100049, China*

[6]*School of Microelectronics, Shanghai University, Shanghai 200444, China*

[#]These authors contributed equally to this work.

[†]jzhao@nudt.edu.cn
[*]jinpingmrg@163.com

**Abstract**

We study population distributions and lasing actions of $N_2^+$ driven by femtosecond lasers with various wavelengths, and uncover an efficient ionic excitation mechanism induced by synchronized ionization and multiphoton resonance. Our results show that the strongest $N_2^+$ lasing appears around 1000 nm pump wavelength. At the optimal wavelength, the pump-energy threshold for air lasing generation is reduced by five folds compared with that required by the previous 800 nm pump laser. Simulations based on the ionization-coupling model indicate that although the Stark-assisted three-photon resonance can be satisfied within a broad pump wavelength range, the optimal pump wavelength arises when the dynamic three-photon resonance temporally synchronizes with the ionization injection. In this case, the ionic dipoles created at each half optical cycle have the same phase. The dipole phase locking promotes the continuous population transfer from ionic ground state to the excited state, giving rise to a dramatic increase of excited-state population. This work provides new insight on the photoexcitation mechanism of ions in strong laser fields, and opens up a route for optimizing ionic radiations.

Photoexcitation and photoionization are the well-known fundamental processes of laser-matter interaction. Photoexcitation of atoms mainly causes the transition of bound electrons from ground state to excited states, while photoionization results in the breaking down of atoms with electrons ejected. In most cases, they are regarded as two independent processes and can be separately dealt with by manipulating the laser parameters. In the strong-field regime, the outermost electrons of atoms or molecules can readily escape from the distorted potential barrier at the instant of light field oscillation [1], and creates extreme non-stationary ionic states triggering attosecond correlated electron-hole dynamics [2-4]. Along with photoionization occurring on the attosecond time scale, photoexcitation of ions can also take place and play an important role. The relevant investigations on the quantum many-body systems have aroused extensive interests of strong-field physics community in the past few decades [5-11].

Recently, some studies have demonstrated the interplay of strong-field ionization and excitation. Sabbar et al. revealed the influence of the ionic polarization on the attosecond dynamics of tunnel ionization by a few-cycle laser pulse [12]. Zhang et al. demonstrated theoretically that the instantaneous ionization breaks the time-reversal symmetry of the light-induced polarization, which enhances the ionic excited-state population [13]. The synergistic effect of photoionization and multiphoton processes can even change the optical radiation properties of bound-bound transitions [14]. Moreover, the advances in $N_2^+$ lasing indicate the excitation process by either polarization or resonant effect is indispensable for inducing a net gain [13, 15-18] since the ionization alone is difficult to cause the population inversion in the ionic system [19, 20]. Very recently, Yuen et al. studied the coherence of tunnel ionization on the population distribution of singly ionized molecules and sequential double ionization [21]. These studies confirm that the photoionization and photoexcitation cannot be considered independently in the strong-field regime.

In this Letter, we study the influence of synchronization between tunnel ionization and multiphoton excitation on the ionic population redistribution, and uncover a novel ionic excitation mechanism. The combined experimental and theoretical results substantialize that the population of ions on the excited state as well as population inversion are maximized upon the tunnel ionization and multiphoton resonance are temporally synchronized. This signifies that the multiphoton excitation efficiency is associated with the birth moment of ions. Our further analysis shows that the time synchronization of ionization and multiphoton resonance results in the phase locking of ionic dipoles created at each half optical period, which allows for continuous population transfer from the lower to the upper levels. These findings shed new light on the multiphoton excitation of ions in ultrafast laser fields, and provide critical guidance for optimizing ionic lasing and hence broadening its applications.

The experiments were performed with a wavelength-tunable pump laser and an external seed with the central wavelength of 391 nm. The schematic of experimental setup is shown in Fig. 1(a). The pump laser was generated by frequency doubling of the idler beam from an optical parametric amplifier, which allows us to continually tune the laser wavelength from 850 to 1070 nm. The seed pulse was provided by frequency doubling of the 800 nm laser. The pump and seed beams were collinearly focused into the nitrogen gas chamber using $f$ = 10 cm lens. The pump energy was measured before the gas chamber. The pump laser intensity was calculated with the assumption of linear propagation, which could be overestimated due to neglecting plasma defocusing. The delay between the two beams was optimized for generating the strongest $N_2^+$ lasing. The polarization directions of pump and seed pulses were perpendicular to each other. The signal was filtered and then was fully collected into a spectrometer for spectral analysis. A polarizer was placed in front of the spectrometer which suppresses the transmission of $N_2^+$ lasing produced by the pump alone.

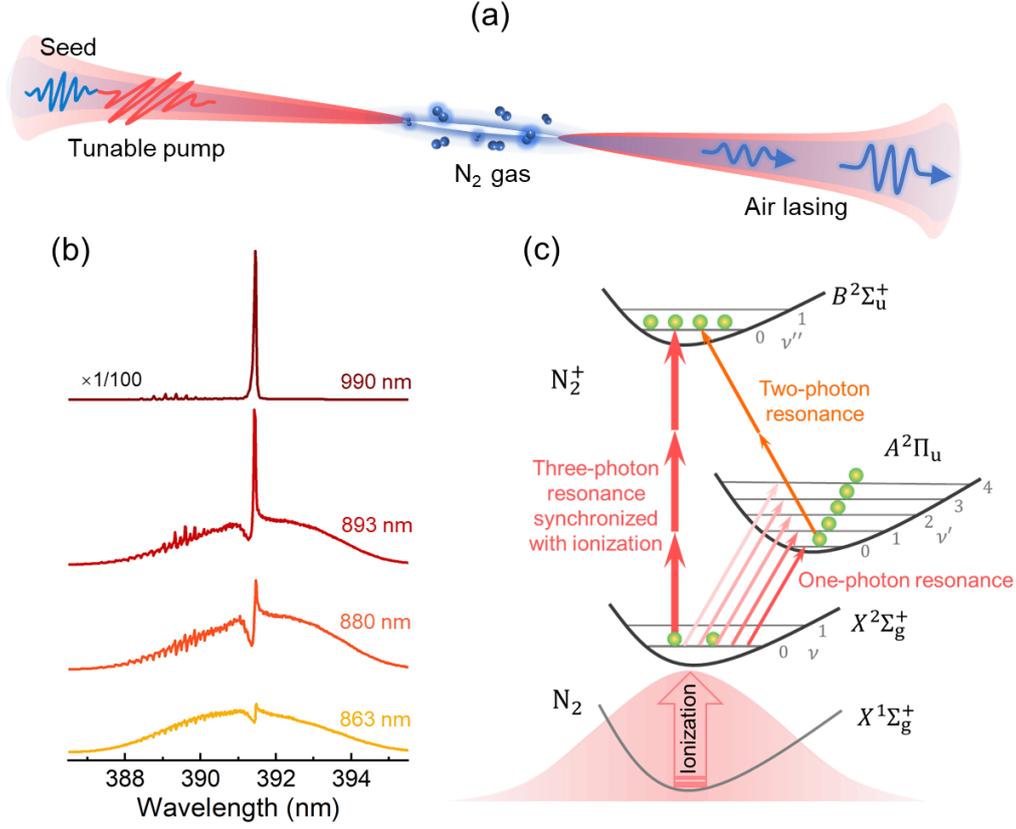

FIG. 1 (a) Schematic of the experiment setup. (b) The typical absorption or emission spectra measured at several pump wavelengths. (c) Schematic illustration of population transfer and redistribution.

In the experiments, the focused intensities of pump lasers at various wavelengths were kept nearly the same by controlling the incident energy, according to the measured pulse durations and focused radii. The minor difference in spatiotemporal characteristics of these pump lasers was ignored. As illustrated in Fig. 1(b), the spectral lineshapes of seed pulses vary from absorption to strong emission while tuning the pump wavelength. Both absorption and emission peaks are located at 391.4 nm, corresponding to the transition between $B^2\Sigma_u^+(v'' = 0)$ and $X^2\Sigma_g^+(v = 0)$. In addition, a remarkable Fano lineshape appears when the pump wavelength is tuned to 880 nm. These results indicate that the population distribution on the two ionic states strongly depends on the pump wavelength. It can be attributed to population transfer between multiple ionic channels in a strong laser field, as shown in Fig. 1(c), which will be elaborated later.

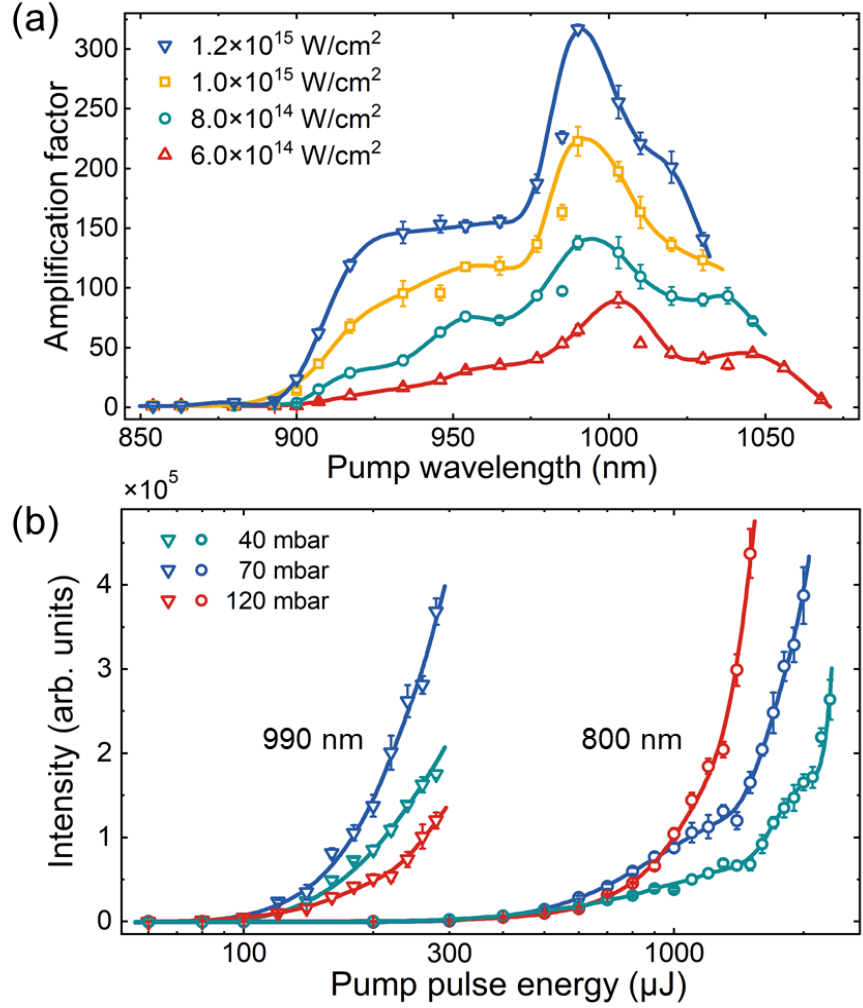

FIG. 2 (a) The amplification factor of seed pulses in the 40 mbar nitrogen gas as a function of the pump wavelength at various pump intensities. (b) The dependence of $N_2^+$ lasing pumped by the 800 nm and 990 nm lasers on the pump energy. Solid curves are numerical fits of experimental data to guide the eyes.

To characterize the amplification capability of the $N_2^+$ systems prepared by various pump lasers, we studied the amplification factor, defined as the ratio of the spectral peak of $N_2^+$ lasing and the spectral intensity of the initial seed at 391.4 nm, as a function of the pump wavelength at several intensities. As shown in Fig. 2(a), the amplification occurs within a broad range from 900 nm to 1050 nm, and the amplification factor gradually grows as the pump intensity increases. For all the measured pump intensities, the maximum gain appears near 1000 nm. When the intensity increases from $6\times10^{14}$ W/cm$^2$ to $8\times10^{14}$ W/cm$^2$, the optimal pump wavelength shifts from 1003 nm to 990 nm. However, the optimal wavelength remains unchanged while further increasing intensity. The maximum amplification factor exceeding 300 is achieved at the pump wavelength of 990 nm and the intensity of $1.2\times10^{15}$ W/cm$^2$.

In previous studies, Ti:sapphire lasers delivering the 800 nm central wavelength were commonly utilized to study $N_2^+$ lasing [15-17,22-25]. Figure 2(b) shows the comparative dependence of $N_2^+$ lasing intensity on the incident laser energy at the two wavelengths of 990 nm and 800 nm. In the measurements, gas pressures were chosen at 40 mbar, 70 mbar and 120 mbar. Note that the focal length of the focused lens was taken as 15 cm for the 800 nm pump laser to ensure nearly the same focal spot size with that of the 990 nm laser. We can clearly see that in the case of 990 nm pumping, the $N_2^+$ lasing begins to generate around 100 μJ, followed by an exponential growth with the increased energy. The pump-energy threshold is reduced to about 1/5 of that with the 800 nm pump laser. At the maximum pump energy of around 300 μJ, the seed can be amplified by 400 times for the 990 nm pumping case, whereas the gain has not yet been established in the 800 nm pumping case. These results demonstrate obvious advantages of the 990 nm pumping for obtaining higher gain and stronger $N_2^+$ lasing.

To interpret the above experimental results, we simulate the evolution of $N_2^+$ population with the following ionization-coupling model [13,14], which is expressed as

$$\frac{d\rho^+}{dt} = -i[H(t),\rho^+] + \left(\frac{d\rho_{nv}^+}{dt}\right)_{ionize} + \left(\frac{d\rho_{nv,n'v'}^+}{dt}\right)_{decay}. \qquad (1)$$

Here, the $\rho^+$ denotes the density matrix of $N_2^+$, $H(t)$ is the Hamiltonian, $n(n')$ represents the $X^2\Sigma_g^+$, $A^2\Pi_u$ and $B^2\Sigma_u^+$ electronic states (simplified as X, A and B), and $v(v')$ represents the vibrational states ($v, v' = 0\sim4$). The terms on the right side of Eq. (1) represent in turn the laser-ion coupling, the transient ionization injection, and the decay process. The laser pulse duration in the simulation is 60 fs according to the experiment, the dephasing time is 1 ps, and the angle between the molecular axis and the laser polarization is assumed as 45°.

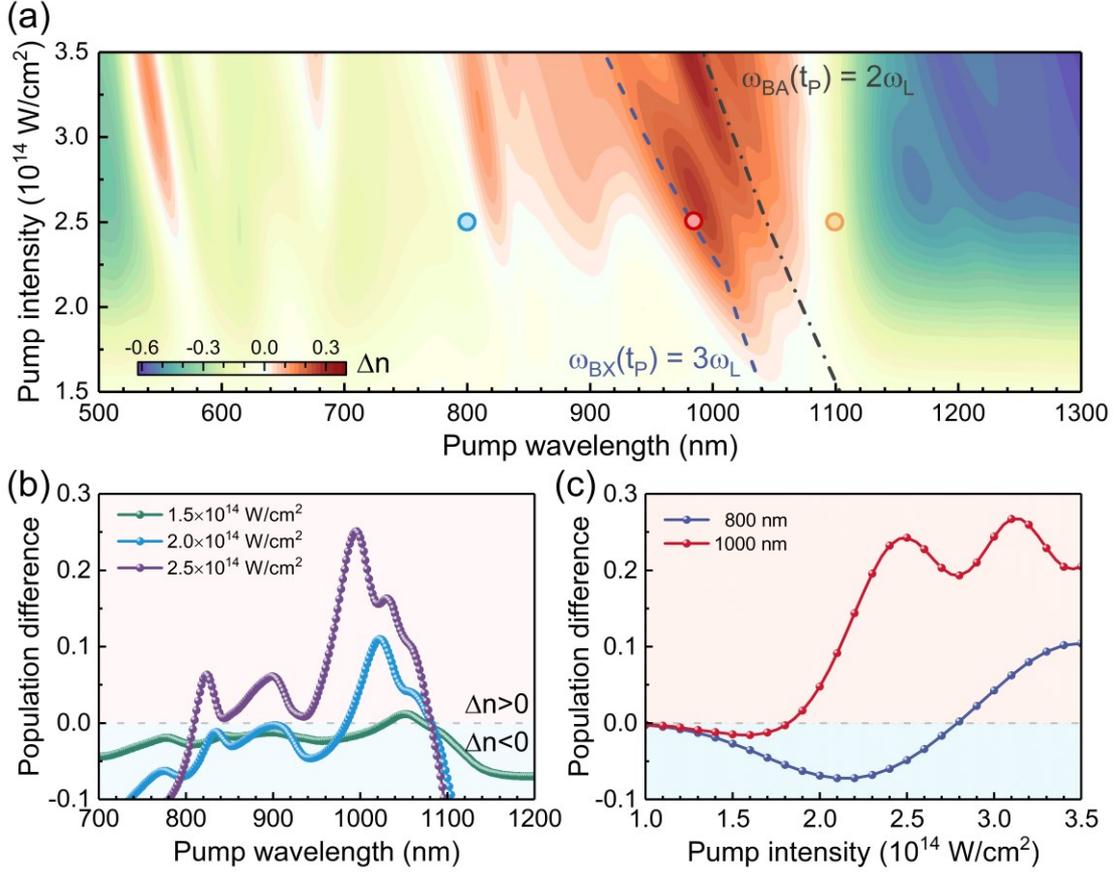

FIG. 3 (a) The calculated population difference between $B(v''=0)$ and $X(v=0)$ states as functions of the wavelength and intensity of the pump laser. (b) The population difference versus the pump wavelength at three typical pump intensities. (c) The dependence of population difference on the pump intensity for the 800 nm and 1000 nm pumping cases.

The intensity of seed-amplified lasing signal at 391.4 nm is mainly determined by the population difference between $B(v''=0)$ and $X(v=0)$ states, i.e., $\Delta n = \rho_{BB}^+ - \rho_{XX}^+$. Figure 3(a) shows the calculated $\Delta n$ as functions of the pump wavelength and intensity. Two strong gain channels basically follow the curves of three-photon and two-photon resonances marked by the dashed and dash-dot lines, indicating the crucial role of the Stark-assisted resonant excitation in population inversion. The minor deviation could originate from the mutual influence of two resonant channels. It is noteworthy that these resonant curves are given when the maximum transition energy of $B(v'' = 0) - X(v = 0)$ and $B(v'' = 0) - A(v' = 0)$ is equal to the three-photon and two-photon energy of the pump laser, i.e., $\omega_{BX}(t_p) = 3\omega_L$ and $\omega_{BA}(t_p) = 2\omega_L$, respectively. Here, $t_p$ is the peak moment of the pump laser field. Therefore, the three-photon resonance mediated by dynamic Stark shift is essentially different from previous studies in which the cycle-averaged transition energy is considered [26-28].

Figure 3(b) plots the quantitative relation of $\Delta n$ with the pump wavelength at three typical intensities. From the simulation results, we can see that the population inversion begins to appear near 1050 nm at the intensity of $1.5\times10^{14}$ W/cm$^2$, and then gradually extends to shorter wavelengths with the increase of pump intensity. When the intensity reaches $2.5\times10^{14}$ W/cm$^2$, the population inversion can be established in a broad range from 800 nm to 1080 nm, and a gain peak appears near 1000 nm with $\Delta n \approx 0.25$. Numerical calculations are in reasonable agreement with our measurements, but quantitative differences still exist owing to the neglect of propagation effects in the simulation. Figure 3(c) compares the pump-energy dependences of $\Delta n$ in the cases of 800 nm and 1000 nm pumping. Apparently, the required pump energy for the establishment of population inversion is much lower for the 1000 nm pumping case, which well explains experimental observations in Fig. 2(b).

Although we reproduce the main experimental results with the calculated population differences, two interesting questions emerge. The first question is how to build up the population inversion in a broad range of pump wavelengths. The other one is that why does dynamic multiphoton resonance cause such high-efficient population transfer. In quest of answers to these questions, we further calculated the evolution of population with the pump wavelength at a given intensity of $2.5\times10^{14}$ W/cm$^2$. As shown in Fig. 4(a), the $X(v = 0)$ population decreases significantly in a broad pump wavelength range. It can be attributed to one-photon excitation from $X(v = 0)$ to multiple vibrational levels of A state (see supplementary) [29]. In addition, the $B(v'' = 0)$ population shows a strong peak around 1000 nm, which approximately matches with the dashed line of the three-photon resonance. Surprisingly, the three-photon excitation is very efficient, even prevails over one-photon excitation from $X(v = 0)$ to $A(v' = 0)$. Hence, the $A(v' = 0)$ population exhibits a dip around the three-photon resonant wavelength.

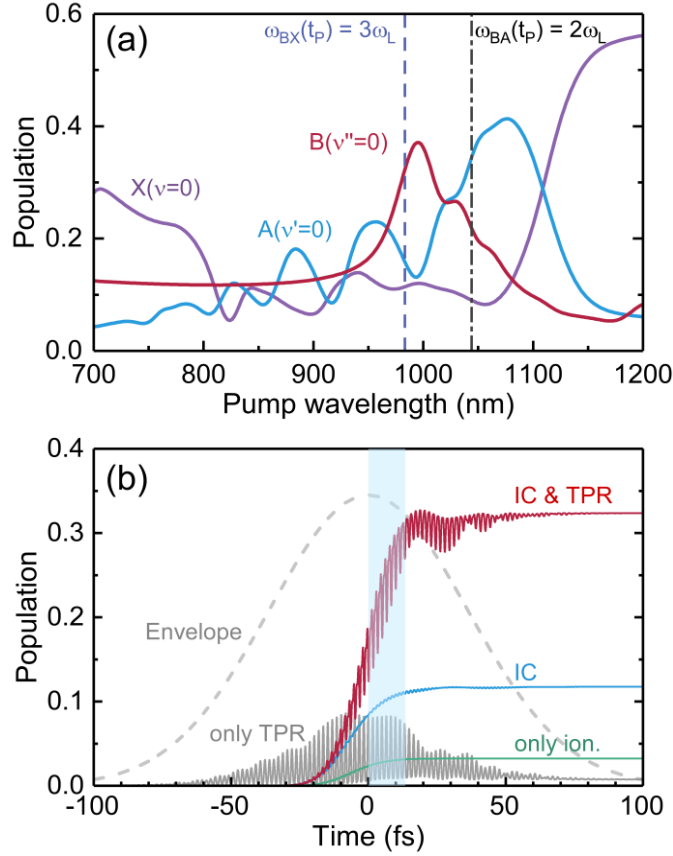

FIG. 4 (a) The population of $X(v = 0)$, $A(v' = 0)$ and $B(v'' = 0)$ states as a function of the pump wavelength. (b) The evolution of population on the $B(v'' = 0)$ state in four cases. The dashed line denotes the electric field envelope. The shaded region indicates the time window of phase analysis in Fig. 5. More details are given in the text.

To gain a deeper understanding for abnormal three-photon excitation, we compared population evolution of the $B(v'' = 0)$ state in four cases, i.e., considering only ionization, only three-photon resonance (TPR), ionization-coupling (IC) with or without TPR, as shown in Fig. 4(b). To compare the results of IC with and without TPR, the pump lasers with the wavelength of 983.3 nm and 800 nm are used, respectively, as indicated by red and blue circles in Fig. 3(a), respectively. Figure 4(b) illustrates that when only considering ionization or TPR, only 3.2% and 0.7% ions are excited to the B($v'' = 0$) state, respectively. In contrast, the simultaneous involvement of ionization and coupling promotes the rapid increase of the B($v'' = 0$) population (blue line) due to instantaneous switch-on of ionic polarization [13]. When the pump wavelength is switched to 983.3 nm, the TPR takes place along with the IC process. In this case, the B($v'' = 0$) population reaches 32.4% (red line), which is 1-2 orders of magnitude higher than that with the TPR only. These results indicate that the transient ionization injection significantly enhances the TPR efficiency.

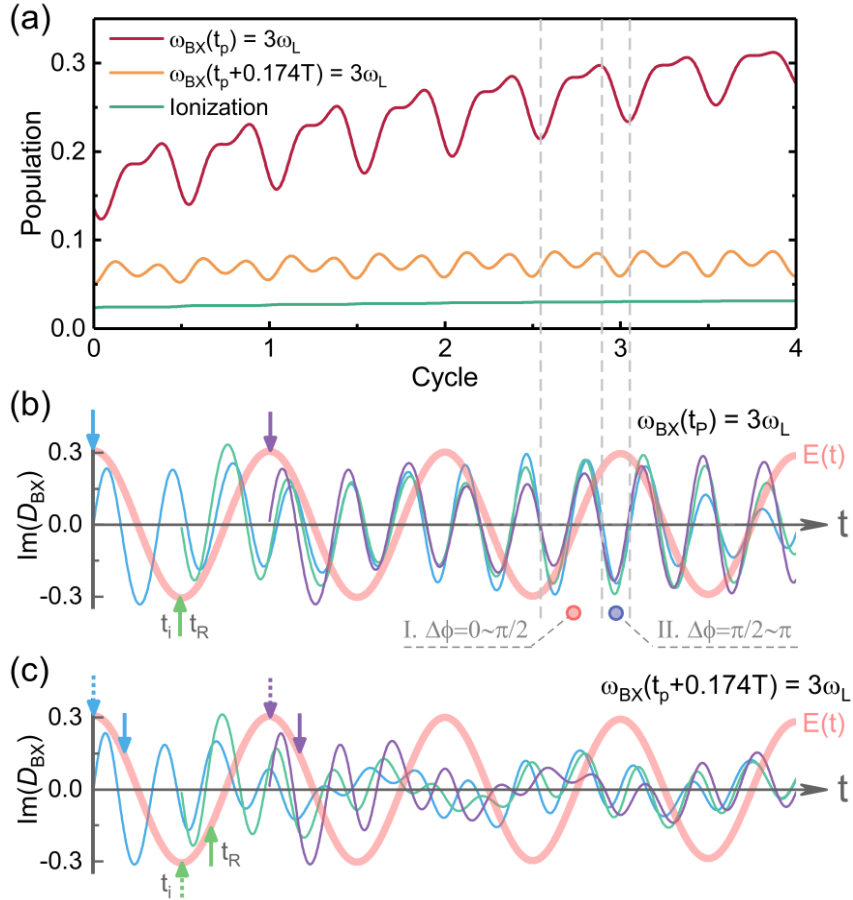

FIG. 5 (a) The population dynamics of the B($v'' = 0$) state calculated by the ionization-coupling equation in the two TPR cases. For comparison, the result calculated with the MO-ADK theory is indicated by green line. The evolution of Im($D_{BX}$) created at three different moments in the case of (b) $\omega_{BX}(t_p) = 3\omega_L$ and (c) $\omega_{BX}(t_p + 0.174T) = 3\omega_L$, in which the TPR is satisfied with the 983.3 nm and 1100 nm pump laser, respectively. The $t_p$ represents the peak moment of the pump laser field. The ionization moment $t_i$ is almost the same as the TPR moment $t_R$ in 5(b), whereas the transient TPR always has a delay with respect to the ionization in 5(c).

To elucidate the influence of ionization injection on the TPR, we consider phase matching between the ionic dipole $D_{BX}$ and the laser field $E(t)$. Since the excitation rate is proportional to $\text{Im}(D_{BX}) \cdot E(t)$ [30], the phase matching between them directly determines the efficiency and direction of population transfer. When their phase difference $\Delta\phi \in [0, \pi/2]$, $N_2^+$ ions will be pumped to the excited state from the ground state via three-photon excitation. The contrary transition will happen for the case of $\Delta\phi \in [\pi/2, \pi]$. Although the Stark-assisted TPR can be satisfied at different moments of the pump laser field by choosing the proper pump wavelength, the phase evolution of the corresponding ionic dipole is quite different. Here, we take the 983.3 nm and 1100 nm pump laser as examples, as indicated by red and yellow circles in Fig. 3(a).

It can be clearly seen from Fig. 5(a) that although the TPR channels are switched on for the two pump wavelengths, the 983.3 nm laser has a higher excitation efficiency in which the TPR takes place at the peak moment of the pump field, i.e., $\omega_{BX}(t_p) = 3\omega_L$. The striking difference stems from different dipole phase evolutions in the two pumping cases, as shown in Fig. 5(b) and (c). Along with the ionization injection, the ionic dipoles are periodically created at each half cycle of the laser field, as shown by the blue, green and purple curves. For the case of $\omega_{BX}(t_p) = 3\omega_L$, the synchronization of ionization and resonant excitation (i.e., $t_i = t_R$) makes the ionic dipole oscillate at the frequency of $3\omega_L$ once it is created. Hence, the dipoles born at different instants possess the same phase. The dipole phase locking promotes the photoexcitation from $X(\nu = 0)$ to $B(\nu'' = 0)$, which can be well understood by the sub-cycle phase analysis. In the region I, $\Delta\phi$ mainly varies in the $0 \sim \pi/2$ range, enabling the increase of excited-state population. In the region II, the contrary process occurs because $\Delta\phi = \pi/2 \sim \pi$ corresponds to the negative excitation rate. With the periodic oscillation of the laser field, the population transfers back and forth between the two states. However, for each half cycle, the population excitation is more efficient than the deexcitation, because in-phase oscillation of $\text{Im}(D_{BX})$ and $E(t)$ lasts longer time. We thus observe the net increase of $B(\nu'' = 0)$ population within each half cycle. The phase locking within multiple optical cycles around the laser field peak ensures the continuous growth of $B(v'' = 0)$ population.

When the pump wavelength is switched to 1100 nm, the TPR is satisfied at the moment of $t_p + 0.174T$. In this case, since the ionization and TPR occur at different moments, i.e., $t_i \neq t_R$, the ionic dipole created at the ionization instant will accumulate a certain phase before the TPR, and the accumulated phase varies as the laser field evolves. Eventually, the ionic dipoles generated at three moments cannot be phase locked, as shown in Fig. 5(c). The phase difference of these dipoles with respect to the laser field also significantly changes for different injection moments. In addition, the population transfer from $X(v = 0)$ to $B(v'' = 0)$ for each half cycle has almost the same efficiency with the contrary process, making the three-photon excitation less efficient. The comparative analysis indicates that the multiphoton excitation efficiency strongly depends on the ionization moment. The synchronization of tunnel ionization and dynamic multiphoton resonance greatly enhances the excited-state population, due to the intrinsic phase locking of dipoles born at different moments. This effect never takes place in neutral atoms or molecules, thus it shows a major advantage of ionic quantum system prepared by strong laser fields.

In summary, we experimentally demonstrated that the optical gain of $N_2^+$ ions can be produced in a broad range of pump laser wavelengths, while the maximum gain is located near 1000 nm rather than 800 nm. The theoretical simulation reproduces the experimental results, and reveals that the excitation efficiency will be dramatically enhanced when the three-photon resonance synchronizes with tunnel ionization. The sub-cycle behavior of multiphoton excitation is attributed to the phase locking of the ionic dipoles generated at different moments, which enables the continuous population transfer from the ionic ground state to the excited state. This work shows a new photoexcitation mechanism, which is universal for ionic systems prepared by ultra-intense ultrafast lasers. The high-efficient excitation scheme opens up great opportunities for generating air lasing by using compact fiber lasers due to the dramatic decrease of the required pump energy.

## Acknowledgements

This work is supported by the National Natural Science Foundation of China (Nos. 12034013, 12234020, 12274428, 12074063, 12264003), Project of Chinese Academy of Sciences for Young Scientists in Basic Research (No. YSBR-042), Program of Shanghai Academic Research Leader (No. 20XD1424200), and Natural Science Foundation of Shanghai (Nos. 22ZR1481600, 23ZR1471700).

# Supplementary Information:

## Enhanced Population on Ionic Excited States by Synchronized Ionization and Multiphoton Resonance

Yewei Chen[1,3,#], Hongbin Lei[2,#], Quanjun Wang[1], Hongqiang Xie[4], He Zhang[1], Xu Lu[1,5], Ning Zhang[1,5], Shunlin Huang[1], Yuzhu Wu[1,6], Jianpeng Liu[2], Qian Zhang[2], Yi Liu[3], Zengxiu Zhao[2], Jing Zhao[2,†], Jinping Yao[1,*]

[1]*State Key Laboratory of High Field Laser Physics, Shanghai Institute of Optics and Fine Mechanics, Chinese Academy of Sciences, Shanghai 201800, China*

[2]*Department of Physics, National University of Defense Technology, Changsha 410073, China*

[3]*School of Optical-Electrical and Computer Engineering, University of Shanghai for Science and Technology, Shanghai 200093, China*

[4]*School of Science, East China University of Technology, Nanchang 330013, China*

[5]*Center of Materials Science and Optoelectronics Engineering, University of Chinese Academy of Sciences, Beijing 100049, China*

[6]*School of Microelectronics, Shanghai University, Shanghai 200444, China*

[#]These authors contributed equally to this work.

[†]jzhao@nudt.edu.cn
[*]jinpingmrg@163.com

**Vibrational population distributions of three electronic states**

We calculated the vibrational population distribution of X, A and B states after the end of the pump laser as a function of the pump wavelength while the pump intensity remaining at $2.5 \times 10^{14}$ W/cm$^2$. As shown in Fig. S1, the population on the X($v = 0$) state is in sequence excited to A($v' = 0, 1, 2, 3, 4$) energy levels via one-photon resonance as the pump wavelength decreases. The population transfer and redistribution processes induced by one-photon resonance are schematically illustrated in Fig. 1(c) of main text. Moreover, it is noteworthy that differing from that of high vibration levels of $v' = 3, 4$ of the A state, the population on the $A(v' = 0, 1, 2)$ state covers a broader range of pump wavelength and overlaps with each other. This could be attributed to large dipole moments between $X(v = 0)$ and $A(v' = 0, 1, 2)$ states and the Raman-type population transfer between these vibrational levels.

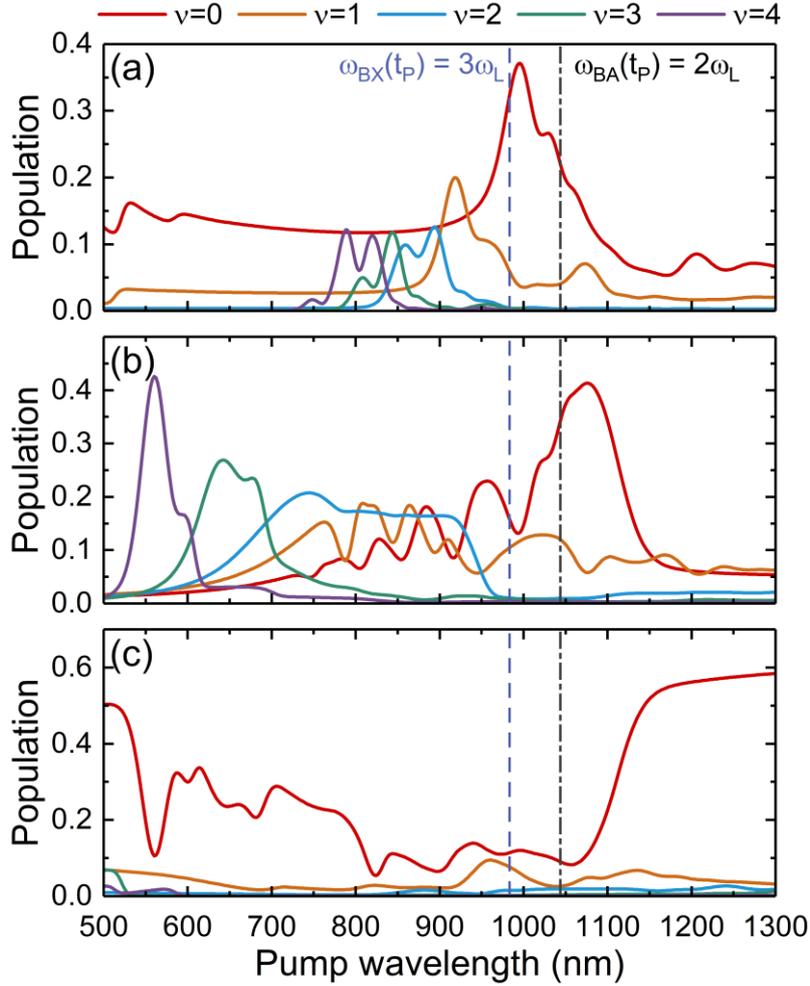

FIG. S1 The vibrational population distribution of the (a) B, (b) A and (c) X states as a function of the pump wavelength. In these simulations, the pump intensity is chosen as $2.5 \times 10^{14}$ W/cm$^2$.

Next, we focus on the population variation of the $B(v'' = 0)$ state. As shown in Fig. S1(a), the population on the $B(v'' = 0)$ state exhibits two peaks around 1000 nm pump wavelength. The large and small peaks basically match with the dashed line of three-photon resonance and the dash-dot line of two-photon resonance, respectively. The minor deviation could originate from the mutual influence of two resonant channels.

The overlapping of the two resonant channels makes them close to each other. Surprisingly, the three-photon resonance from $X(v = 0)$ to $B(v'' = 0)$ is very efficient, even prevails over one-photon resonance from $X(v = 0)$ to $A(v' = 0)$, resulting in a dip for the population of the $A(v' = 0)$ state around 1000 nm. In addition, the transient polarization of the ionic system can cause a certain population on the $B(v'' = 0)$ state, which is insensitive to the pump wavelength. As a consequence, the population on the $B(v'' = 0)$ state exhibits a flat distribution beyond the region of three-photon resonance, as illustrated in Fig. S1(a). Other vibrational levels of the B state can also be efficiently populated via similar resonant excitation when the pump laser is tuned to proper wavelengths.

The vibrational-resolved results reveal that (i) the broadband gain mainly results from the decrease of $X(v = 0)$ population via multi-channel one-photon resonance and the increase of $B(v'' = 0)$ population caused by ionic transient polarization; (ii) the gain peak is attributed to the significant increase of $B(v'' = 0)$ population via dynamic-Stark-assisted three-photon resonance. Unlike the conventional multiphoton resonance, the three-photon excitation here has an extremely high efficiency due to the synchronization of ionization and multiphoton resonance, as demonstrated in the main text.